\documentclass[sigconf]{acmart}

\renewcommand\footnotetextcopyrightpermission[1]{} % removes footnote with conference information in first column

\setcopyright{none}

\acmConference[VEM 2026]{14th Workshop on Software Visualization, Maintenance and Evolution}{September 11, 2026}{São Paulo, SP, Brazil}

\AtBeginDocument{
    
}

\begin{document}

%% allowing the author to define a "short title" to be used on the page 
%% headers.
\title{Exploring Semantic Stability Across Reviews in the Linux Kernel}
%\title{A First Look at Semantic Stability Across Reviews in the Linux Kernel}
%\title{A First Look at Semantic Stability Across Reviews in the Linux IIO Subsystem}

\author{Lucas Ciziks}
\affiliation{
  \institution{Universidade de São Paulo (IME-USP)}
  \city{São Paulo}
  \country{Brazil}
}
\email{luciziks@usp.br}

\author{Paulo Meirelles}
\affiliation{
  \institution{Universidade de São Paulo (IME-USP)}
  \city{São Paulo}
  \country{Brazil}
}
\email{paulormm@ime.usp.br}

\author{Marco Aurélio Gerosa}
\affiliation{
  %\institution{School of Informatics, Computing, and Cyber Systems}
  \institution{Northern Arizona University (NAU)}
  \city{Flagstaff, AZ}
  \country{USA}
}
\email{Marco.Gerosa@nau.edu}

%% By default, the full list of authors will be used on the page
%% headers. This list is often too long and will overlap
%% other information printed in the page headers. 
%% This command allows the author to define a more concise list
\renewcommand{\shortauthors}{Ciziks et al.}
\renewcommand{\shorttitle}{Exploring Semantic Stability in the Linux Kernel}

%% The abstract is a short summary of the work the paper presents.
\begin{abstract}
Code review is credited with substantially changing a patch's code between its first submission and the version that eventually lands. However, prior work typically studied only the final merged patch without comparing it to the first submission. We present a function-level measurement that tracks 10,117 trajectories (each function followed across the numbered revisions of one patch series) through the patch history of the Linux IIO subsystem, comparing similarity scores against unrelated function pairs as a 
baseline. A naive reading yields near-total similarity, but this is largely an artifact of composition: 75.3\% of tracked trajectories are never textually modified between versions, contributing a trivial 100\% similarity that inflates the headline. Restricting to the trajectories with a real edit, semantic 
purpose is still largely preserved (mean similarity 0.990 vs.\ a 0.909 
baseline), but drift appears to concentrate in the first review round mainly 
because later rounds contain more functions that nobody touched, not 
because edits become more conservative over time. After controlling for it, a statistically detectable but small residual effect remains. This points to an open question: whether near-ceiling similarity reflects preserved purpose or a measurement 
tool that cannot detect the significance of small, localized edits. We present this work as a first look and outline next steps.
\end{abstract}

%% the presented work. Separate the keywords with commas.

%% This command processes the author, affiliation, and title
%% "frenchspacing" avoids an additional space after a period at the end of a sentence.
\frenchspacing
\maketitle

\vspace{-.2cm}
\section{Introduction}
\label{sec:introduction}

A long review thread on a kernel mailing list is often read as evidence that a contribution was contested. A substantial body of patch analysis work nonetheless treats the final merged commit as the primary unit of analysis~\cite{patchnet, islam2026, nguyen2025}, leaving aside how fixes evolve before landing. A recent large-scale study of Linux kernel bug-fix lifecycles finds that accepted repairs frequently spread across files and functions beyond where the fault was first observed, as reviewer feedback enforces constraints not visible in the initial bug report~\cite{beyondcrash}. What is missing is a direct measurement of the semantic side of that evolution: does review change what a function is \emph{for}, or does it converge on an implementation of a purpose already fixed?

We address this question in the Linux IIO (Industrial I/O) subsystem, where our own contributions to IIO drivers gave us repeated first-hand exposure to the phenomenon: v1 and vN code differ substantially line by line, even though reviewers and authors treat the series as converging on ``the same fix.'' This gap between visible diffs and perceived semantic stability motivates our study. In the kernel's submission convention~\cite{kernel-submitting-patches}, a patch series progresses through numbered revisions (v1, v2, \ldots, vN) on the mailing list before being merged or abandoned; we track each function across these revisions and pose two research questions:

\textbf{RQ1.} \textit{Does a function's semantic purpose change between the first and last submitted version of a patch series, relative to what unrelated code looks like in the same embedding space?}

\textbf{RQ2.} \textit{If some drift occurs, is it spread evenly across revisions, or concentrated in a specific part of the process?}

This distinction matters beyond the IIO subsystem. Tools that score a patch averaging similarity across all functions in a series, as in patch classification, fix-to-bug matching, or automated review assistants, risk treating ``no edit occurred'' and ``review preserved purpose'' as the same signal. As we show below, a large share of tracked functions in a patch series are never retouched after their first submission, contributing a similarity score of 1.0 to that aggregate. A near-ceiling score is consequently a weak guarantee: it can reflect a genuinely stable function, or simply one no reviewer touched, or a small, localized edit the embedding is not sensitive enough to register. Left unaddressed, this conflation can make semantic similarity metrics appear more reliable than they are, particularly for small, targeted edits, such as security fixes that matter most. We argue that disaggregating by whether an edit actually occurred should be a baseline requirement for this kind of measurement, not an optional refinement.

Our main contribution is a \textbf{research perspective}: a function-level measurement of semantic stability across kernel patch review, with an account of what a naive reading of that measurement gets wrong and why. We also identify a limitation in whole-function cosine similarity over pretrained code embeddings and lay out a research agenda.

\section{Related Work}
\label{sec:relatedwork}
 
Prior patch analysis work, including kernel-specific classifiers such as 
PatchNet~\cite{patchnet} and broader VFC identification approaches~\cite{islam2026, 
nguyen2025}, operates on the landed form of a patch, leaving open whether 
that form is semantically representative of the fix throughout its review 
history. Our RQ1 addresses this gap directly. The closest related work is 
a large-scale study of Linux kernel bug-fix lifecycles~\cite{beyondcrash}, 
which finds that accepted repairs frequently spread across files and functions 
beyond where the fault was first observed, as reviewer feedback enforces 
constraints not visible in the initial report. Tracking a patch through its 
pre-integration history on a mailing list is a known hard problem. Ramsauer 
et al.~\cite{ramsauer2019} address it with a language-independent similarity 
method validated against a hand-built ground truth. We use a coarser, cheaper 
key (normalized subject line, file path, function name), adequate for 
within-series tracking but a natural point of comparison for follow-up work. 
As we note in Section~\ref{sec:agenda}, we plan to quantify its error rate 
against that ground truth.

Our data comes from the LKML5Ws dataset~\cite{passos2026lkml5ws} and complements DUKS~\cite{duks2025vissoft}, a dashboard that unifies mailing-list and git-tree data to visualize kernel process-level evolution, including contributor activity and commit flow between trees. That dataset lacks a code-content dimension. A validated semantic-stability metric like ours would be a natural addition.

The issue we discuss in this paper is also documented elsewhere. Raw cosine similarity in transformer embedding spaces is known to be inflated by anisotropy, so even unrelated pairs score artificially high~\cite{ethayarajh2019}. Pretrained code encoders such as CodeBERT~\cite{codebert} and UniXcoder~\cite{unixcoder} share this architecture, so the same risk plausibly applies to them. We control for this with a random-pair null throughout. Nevertheless, two recent results suggest this correction is not enough. Nikiema et al.~\cite{nikiema2025} systematically tested 18 similarity measures under controlled small transformations and found that embedding-based methods, cosine similarity in particular, routinely score semantically opposite code as similar, while switching from cosine to Euclidean distance on the same embeddings improves discrimination by 24 to 66\%. Farhad and Dass~\cite{farhaddass2026} compare vulnerable and patched function pairs across multiple encoders, including UniXcoder, and find similarity near 0.99 for nearly all pairs across models regardless of whether the patch closes a real vulnerability. They interpret this not as encoder failure but as evidence that whole-snippet semantic similarity cannot detect relevant, spatially localized changes, and complement it with a structural (AST-based) signal. Both, in different corpora, describe the same problem we report below.

\section{Data and Method}
\label{sec:data}

% The pipeline proceeds in four stages: corpus assembly, trajectory construction, embedding, and corpus characterization. The last stage (\emph{Composition of the sample} below) characterizes what the corpus is made of and informs how to read the similarity numbers reported in Section~\ref{sec:results}.
% \vspace{-.15cm}

\paragraph{\textbf{Corpus.}} We build on LKML5Ws~\cite{passos2026lkml5ws}, a dataset of Linux Kernel Mailing List messages, and extract function-level records for the IIO subsystem. For each patch, we fetch the pre-patch file blob via \texttt{git cat-file}, reconstruct a standalone single-file diff, and apply it with \texttt{git apply} in an isolated scratch directory to obtain the real post-patch file. Rows are dropped when (1)~the blob SHA is absent from the local tree, (2)~\texttt{git apply} fails because the hunk context no longer matches the fetched blob (e.g., the email's diff was generated against a slightly different base), (3)~the file is newly introduced or deleted, (4)~the file extension is outside \texttt{\{.c, .h, .rs\}}, or (5)~the touched function cannot be matched unambiguously by name in both the before- and after-file. This conservative strategy yields 64,044 function-version records.
\vspace{-.25cm}

\paragraph{\textbf{Trajectories.}} A trajectory is one function followed across the versions of one series, keyed by (normalized subject line, file path, function name), a key that remains stable across v1..vN, unlike the per-email message ID. Of 37,643 trajectories, 10,117 have at least two clean versions and therefore a defined endpoint pair $(v_1, v_N)$. Of those, 4,387 have three or more. This key is a heuristic rather than a validated identity-resolution method, such as that of Ramsauer et al.~\cite{ramsauer2019}. Quantifying its mislinking or false-drop rate on this corpus is part of our research agenda (Section~\ref{sec:agenda}).
\vspace{-.25cm}

\paragraph{\textbf{Embedding.}} Each function body, before and after a patch, is embedded with UniXcoder~\cite{unixcoder} (\texttt{microsoft/unixcoder-base}), using attention-mask-weighted mean pooling over the last hidden state~\cite{sentencebert}, truncated to the model's 512-token pretraining limit. Functions in this corpus average $\approx$40 LOC, so truncation affects only the long tail; the number of truncated inputs is logged per run. Every distribution below is reported alongside a random-pair null, the similarity of embeddings for unrelated function pairs sampled from the same corpus (mean 0.910). An observed similarity is informative only to the extent it clears this floor; though, clearing it turned out to be a weaker guarantee than we initially assumed.
\vspace{-.25cm}

\paragraph{\textbf{Data Quality.}} We validate function-body reconstruction directly with two checks, rather than assuming hunk-based extraction gets boundaries right by construction. First, a signature-presence check flags trajectories in which a function introduced partway through a series was attached to an unrelated code fragment. Second, we test whether every reconstructed function body is brace-balanced (with equal opening and closing braces) and string-literal-aware. This approach revealed a boundary-detection issue in an earlier version of the extractor, which we corrected before computing the results by re-embedding the full corpus against brace-balanced boundaries. In the corrected corpus, 96.5\% of before- and after-bodies are brace-balanced. The remainder is attributable to macro-heavy IIO code whose braces are unbalanced inside string or preprocessor content for legitimate syntactic reasons.
\vspace{-.25cm}

\paragraph{\textbf{Composition of the Sample.}} Before trusting any similarity number, we characterize the sample by comparing the \emph{literal function-body text}---not the embedding---across consecutive and endpoint versions. Of the 10,117 endpoint trajectories, 7,616 (75.3\%) have \emph{byte-identical} $v_1$ and $v_N$ bodies: the function appears in the diff context of a later revision but is never re-touched after its first submission. Hence, the encoder receives identical text for $v_1$ and $v_N$ and reports a similarity of exactly 1.0, conveying no information about whether the review preserves its purpose. The remaining 2,501 trajectories (24.7\%) carry a real edit between $v_1$ and $v_N$. Restricted to this subset, mean endpoint similarity is 0.990 (median 0.992, 25th percentile 0.989, minimum 0.748), still above the 0.909 null for this subset (Mann-Whitney $U=6{,}206{,}684$, $p\approx0$), but lower and more spread out than the pooled headline figure of 0.997, which includes the trivial cases. The same pattern holds at the transition level: of 18,656 consecutive-version transitions, 15,489 (83.0\%) involve no textual change to the tracked function that round.
\vspace{-.25cm}

\paragraph{\textbf{Measurements.}} We compute three quantities from the same embedding cache. \emph{M1} is a sanity check: cosine similarity between the before and after version of a function within a single patch version, binned by lines changed, confirming the embedding tracks edit magnitude. \emph{M2}, endpoint drift, is the cosine similarity between $v_1$ and $v_N$ for each trajectory with at least two clean versions, the direct measurement for RQ1. \emph{M3}, consecutive-transition drift, is the cosine similarity between adjacent clean versions ($v_i$, $v_{i+1}$), grouped by transition index, the direct measurement for RQ2. All comparisons use the Mann-Whitney U test~\cite{mannwhitney1947}, since similarity scores are strongly left-skewed rather than normally distributed. Given the small effect sizes found (Section~\ref{sec:results}), $p$-values are treated as indicative rather than confirmatory and we rely on effect size where available.

\section{Results}
\label{sec:results}

%TODO: aqui também, se der, é bom para uma visao geral, antes de entrar em cada ponto, para não ir direto.
We report results in the order of Section~\ref{sec:data}: M1 (sanity check), M2 (endpoint drift, RQ1), and M3 (consecutive-transition drift, RQ2).

\begin{figure}[ht]
\vspace{-.4cm}
\centering
\includegraphics[width=\linewidth]{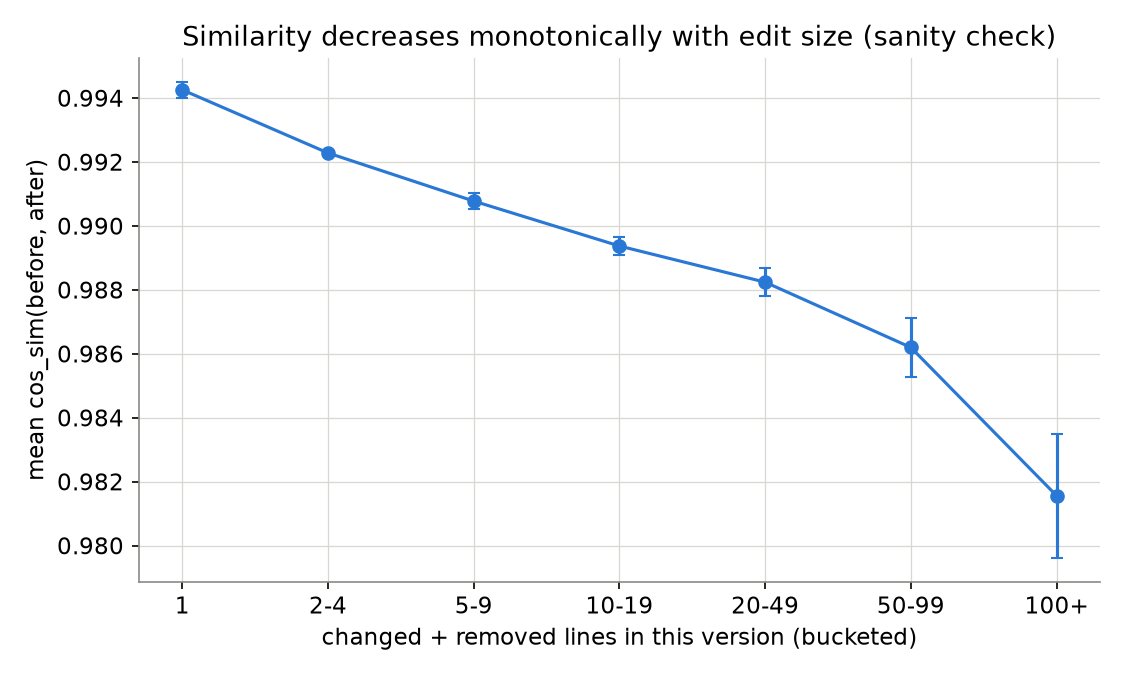}
\vspace{-.7cm}
\caption{Mean within-version similarity by edit size, with 95\% confidence intervals (M1, $n=64{,}044$). Similarity decreases monotonically as edits increase, but remains well above the 0.910 null even in the largest bucket.}
\Description{Line chart with error bars showing mean cosine similarity between a function before and after version within a single patch, grouped by buckets of changed plus removed lines from 1 to 100 or more. The mean declines steadily from about 0.994 at a single changed line to about 0.982 at 100 or more changed lines.}
\label{fig:editsize}
\vspace{-.4cm}
\end{figure}

\paragraph{\textbf{M1: sanity check.}} Mean within-version similarity is 0.991 against a null of 0.910, and decreases monotonically as edits get larger (Figure~\ref{fig:editsize}), from 0.994 at a single changed line to 0.982 above 100 lines. The embedding responds to edit size in the expected direction. Even the largest bucket (100+ lines) sits at 0.982, above the 0.910 null.

\begin{figure}[ht]
\centering
\includegraphics[width=0.85\linewidth]{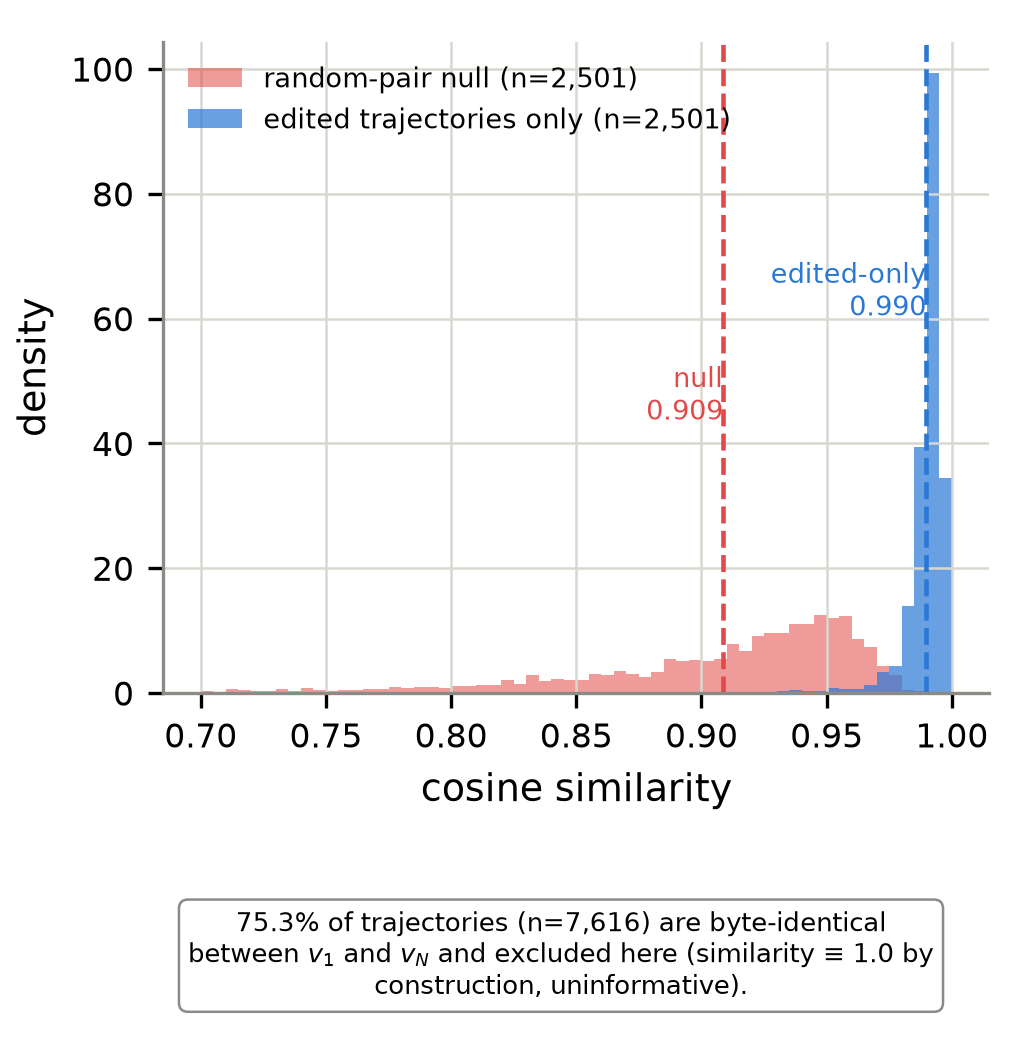}
\vspace{-.5cm}
\caption{Distribution of $\cos(v_1, v_N)$ restricted to the 2,501 trajectories with a real edit between $v_1$ and $v_N$ (mean 0.990), against a random-pair null of the same size (mean 0.909). All percentages shown derive from byte-level textual comparison, not from the embedding.}
\Description{Histogram comparing the distribution of cosine similarity for the 2,501 trajectories with a real edit against a matched random-pair null of the same size. The edited-trajectories distribution is concentrated near 0.99, while the null baseline is spread lower, around 0.91. A note in the figure states that the 7,616 byte-identical trajectories are excluded because unchanged text is guaranteed to produce a similarity of 1.0, which is not a meaningful measurement.}
\label{fig:endpoint}
\vspace{-0.5cm}
\end{figure}

\paragraph{\textbf{M2: endpoint drift (RQ1).}} Across 10,117 trajectories, mean cosine similarity between $v_1$ and $v_N$ is 0.997 (median 1.000, std 0.007, minimum 0.748), against a null mean of 0.910 (Mann-Whitney $U=102{,}154{,}592$, $p\approx0$). This result answers RQ1: the semantic purpose is preserved end-to-end for almost every trajectory. However, three-quarters of that mass is trivial: the underlying text never changed, yielding a similarity of 1.0. Figure~\ref{fig:endpoint} shows the distribution restricted to the 2,501 trajectories with a real edit instead: a still-high but more modest mean of 0.990 against a matched random-pair null of the same size (mean 0.909).

A separate concern is whether 0.990 merely restates that most tracked functions receive small cumulative edits: trajectories with 500+ cumulative edited lines (1.3\% of all 10,117 trajectories; $n=134$) still average 0.996 similarity, and even the 1000+ bucket ($n=10$) averages 0.993, far above the 0.910 null. Cumulative edit volume moves the number, but not enough to explain the ceiling effect.

\vspace{-.25cm}

\begin{figure}[ht]
\centering
\includegraphics[width=\linewidth]{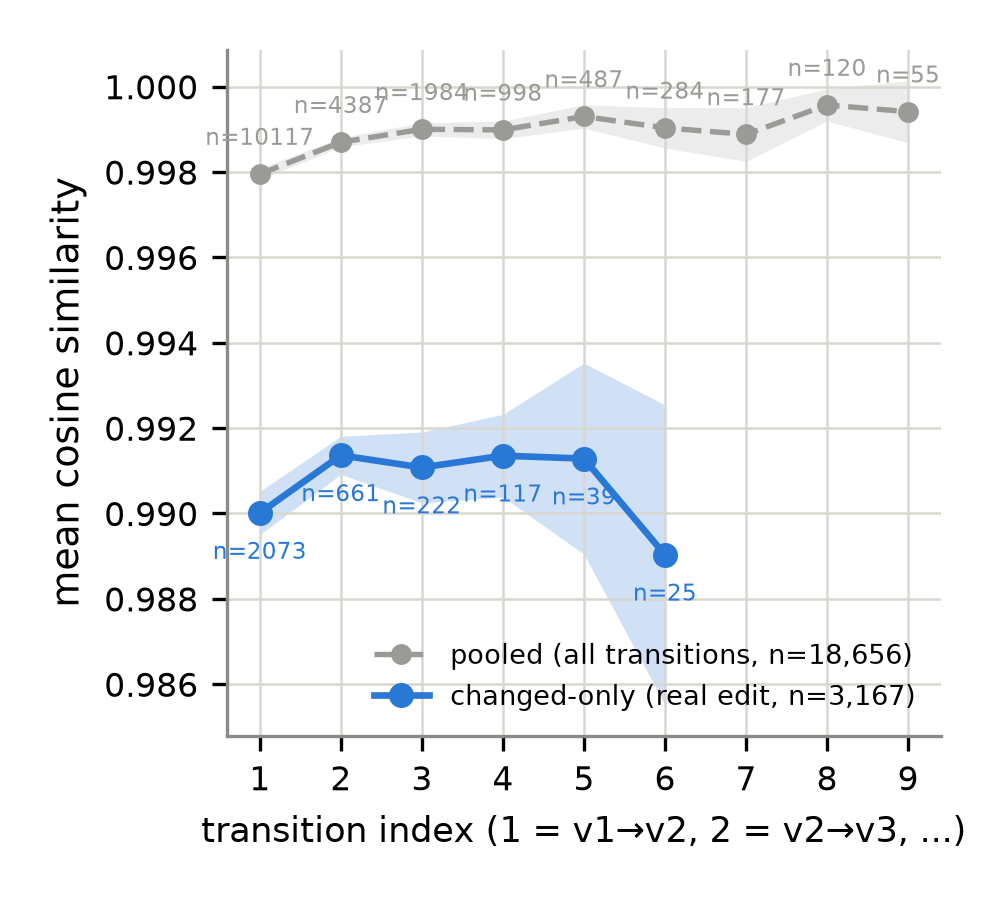}
\vspace{-.7cm}
\caption{Mean similarity by ordinal transition index: pooled (all transitions, gray dashed) versus restricted to transitions with a real textual edit that round (blue solid). The changed-only series is plotted only through the sixth transition; beyond that point, fewer than 25 genuinely edited transitions remain and per-index estimates are unreliable.}
\Description{Line chart comparing mean cosine similarity by transition index for two series: a pooled series including all transitions, shown as a gray dashed line staying near 0.998 to 1.000 across nine transitions, and a changed-only series restricted to transitions with a real textual edit, shown as a blue solid line hovering around 0.989 to 0.991 with a widening confidence band, plotted only through the sixth transition where the sample size drops to 25.}
\label{fig:transitions}
\end{figure}

\vspace{-0.5cm}
\paragraph{\textbf{M3: consecutive-transition drift (RQ2).}} Across 18,656 transitions, mean similarity in the pooled series (Figure~\ref{fig:transitions}, gray dashed line) rises from the first to the fifth transition (0.99795 to 0.99930) and becomes noisier at higher indices, where the sample size per transition is small. Pooling the first transition against all later ones, the first round has lower similarity (mean 0.998) than later rounds (mean 0.999, Mann-Whitney $U=40{,}164{,}232$, $p=1.12\times10^{-18}$). This difference is statistically real, but on a scale ($10^{-3}$) small relative to the full 0-to-1 range of cosine similarity.

This pooled result has an unanticipated confound: the fraction of transitions with no textual change rises, non-monotonically, from 79.5\% at the first transition to a range of roughly 88--95\% among later transitions, peaking at 95.0\% by the eighth. If later rounds contain more trivial similarity-1 cases, that alone would make them appear more stable without genuine edits becoming more conservative. Restricting to transitions where the function's text changed that round (3,167 of 18,656, 17.0\%) reduces the effect (Figure~\ref{fig:transitions}, blue solid line): first-transition mean similarity is 0.9900 (median 0.9923, $n=2{,}073$) versus 0.9912 (median 0.9927, $n=1{,}094$) for later transitions, still in the same direction but far weaker (Mann-Whitney $U=1{,}073{,}784$, $p=6.98\times10^{-3}$ vs.\ $p=1.12\times10^{-18}$ unfiltered). Plotting the changed-only series at each transition makes the fragility of this comparison visible: the sample size falls from 2,073 real edits at the first transition to 25 by the sixth, and the confidence band widens accordingly. We stop plotting beyond that point because fewer than 25 genuinely edited transitions remain, and a per-index estimate would be meaningless. To assess how much this difference matters in practice, we compute its rank-biserial correlation~\cite{cureton1956}, a standard effect-size measure. The result is small ($r\approx0.05$ on a $-1$ to $1$ scale): the two groups' similarity distributions overlap almost completely despite the formal statistical significance. Therefore, most of what appears as review converging over rounds in the pooled M3 result is better explained by later rounds touching the function less often. The residual signal - that edits become marginally more conservative across later rounds - is statistically significant but practically small.

\vspace{-.15cm}
\section{Discussion}
\label{sec:discussion}
 
M2 and M3 support a specific claim: \textbf{review overwhelmingly preserves a function's semantic purpose, and the little drift that exists concentrates in the first round.} This claim requires two qualifications. First, 75.3\% of endpoint trajectories and 83.0\% of consecutive transitions involve no textual change. The embedding cannot detect drift where none exists, inflating both the M2 headline and the apparent M3 convergence pattern. Second, once restricted to trajectories and transitions with a real edit, the RQ1 result holds (0.990 vs.\ a 0.909 null). However, the RQ2 answer changes substantially: the apparent concentration of drift in the first round is substantially accounted for by the rising rate of untouched transitions across the series, though a residual effect remains after controlling for it. What remains is a weaker, small-effect-size version of the same pattern ($r\approx0.05$). This paper's most concrete contribution is to locate and quantify the extent to which a reported effect (RQ2) was an artifact of sampling composition.

A residual concern applies to the edited-only M2 subset: even heavily edited trajectories (500+ cumulative lines, $n=134$) still average 0.996 similarity, far above the 0.910 null. The likely mechanism is dilution: mean pooling across all token representations produces a single 768-dimensional vector, redistributing the signal of a localized 2-line fix across hundreds of token representations that did not change, leaving the resulting vector nearly identical to its predecessor. The weak but real $\rho=-0.146$ correlation with edit volume confirms the model is sensitive to \emph{some} of this signal---but the ceiling persists because localized change is diluted. This is the resolution problem: the instrument cannot distinguish a semantically neutral whitespace fix from a critical concurrency correction if both touch a similar number of tokens inside a large unchanged function. This structural insufficiency matches what Nikiema et al.~\cite{nikiema2025} and Farhad and Dass~\cite{farhaddass2026} independently report in other corpora: \textbf{cosine similarity on whole-snippet embeddings is structurally the wrong resolution for spatially localized relevant change.} Whether near-ceiling similarity on genuinely edited functions reflects preserved purpose or this resolution limit remains the open question we cannot answer from whole-function embeddings alone.

% A domain-specific hypothesis is worth naming: IIO drivers largely wrap a small, framework-defined callback interface, and much review feedback targets locking, error handling, or sysfs conventions rather than what a function computes. If this generalizes, both halves of our composition finding could share a common explanation: untouched neighboring functions across a driver file, and shallow rather than semantically deep edits. We flag this as a hypothesis, not a finding: distinguishing it from a subsystem-independent effect is why the corpus extension in Section~\ref{sec:agenda} matters.

\section{A Research Agenda}
\label{sec:agenda}

We propose the following next steps:

\textbf{Report the edited-only subset by default, not only the aggregate one.} Prior work that scores patches from a single snapshot~\cite{patchnet} does not distinguish a real edit from an untouched function. We show this distinction is not cosmetic: the untouched fraction rises across review rounds (79.5\% at the first transition, higher at later ones), and pooling it in produces the apparent semantic convergence in M3; controlling for it reduces but does not eliminate the effect.

\textbf{Validate the trajectory-linking heuristic and sharpen the instrument.} A stratified annotation protocol has been prepared: a blinded annotation sheet covering 100+ trajectories sampled across five endpoint-similarity strata---including the ten highest-edit-volume cases---is ready for human review. Executing this annotation and reporting the mislinking rate against the validated method of Ramsauer et al.~\cite{ramsauer2019} is the immediate next step. In parallel, we plan to re-run M2 and M3 with Euclidean distance on the same UniXcoder vectors, following Nikiema et al.~\cite{nikiema2025}, and to embed the diff region itself~\cite{cc2vec2020} rather than the whole function, which should be more sensitive to a small edit diluted inside a large unchanged context.

\textbf{Add a structural signal and validate by hand.} Following Farhad and Dass~\cite{farhaddass2026}, pairing the semantic score with an AST edit-distance signal would flag cases where the two disagree, such as a small textual edit with a large structural change, or vice versa. No automated metric substitutes for a kernel-literate reviewer confirming, on a sample of small edits, high-similarity cases, whether the edit is cosmetic or has an outsized behavioral effect, the only way to calibrate what a similarity threshold means for this corpus.

We also plan to extend the corpus beyond IIO to test whether the composition pattern reported here is specific to this subsystem or holds for kernel patch review generally.

\vspace{-.15cm}
\section{Final Remarks}
\label{sec:conclusion}

This paper is a first step toward a broader research perspective: measuring, rather than assuming, how much semantic purpose a function retains across a kernel patch series, at a resolution that matters, since most of what reviewers touch is a handful of lines embedded in an otherwise stable function. That resolution is the central problem. Tracking function-level trajectories through the Linux IIO subsystem with whole-function embeddings, we show that a naive, aggregate reading of semantic similarity is not trustworthy on its own: once similarity is measured against a proper null and disaggregated by how much a function was touched, the instrument's present resolution is not fine enough to reliably separate small but meaningful edits from noise. This limitation is not specific to our corpus or encoder. Independent work on vulnerability patches and controlled code transformations encounters the same limitation, which is why we treat it as a research opportunity rather than a solution.

Section~\ref{sec:agenda} lays out the next steps for this direction: validating how trajectories are tracked across a series, embedding the edit rather than the whole function, pairing semantic and structural signals, and grounding similarity thresholds in human judgment. Our contribution is a \textbf{working measurement pipeline for this question, an account of where it currently falls short, and an initial basis for a line of research on semantic stability in patch review}.
 
\section*{Artifact Availability}
 
The extraction pipeline, embedding scripts, derived metric tables, and analysis notebook used to produce the results in this paper are available at \url{https://doi.org/10.5281/zenodo.21853709}.

\normalsize
\bibliographystyle{ACM-Reference-Format}
\bibliography{references}

%% If your work has appendices, this is the place to put them
% \appendix

\end{document}